
\magnification=1200
\baselineskip=18pt plus 3pt minus 1pt
\hsize=6 truein
\def\ck{c_{\bf k}}
\null
\hfill {KFKI-RMKI-16-FEB-1993}
\vskip 4cm
\centerline{\bf K\'arolyh\'azy's quantum space-time generates}
\centerline{\bf neutron star density in vacuum}
\vskip 1cm
\centerline{Lajos Di\'osi\footnote{$^1$}{diosi@rmki.kfki.hu}
      and B\'ela Luk\'acs\footnote{$^2$}{lukacs@rmki.kfki.hu}}
\centerline{KFKI Research Institute for Particle and Nuclear Physics}
\centerline{H-1525 Budapest 114, POB 49, Hungary}
\vskip 4cm
\line{\bf Abstract}

By simple arguments, we have shown that K\'arolyh\'azy's
model overestimates the
quantum uncertainty of the space-time geometry and leads to
absurd physical consequences. The given model can thus not account
for gradual violation of quantum coherence and can not predict tiny
experimental effects either.
\vfill
\eject

In a pioneering paper [1], it was suggested that the
quantum mechanics of macroscopic objects ought to  be modified
due to a certain eventual unsharpness of space-time
geometry. Later on, the possibility of experimental
verification of the model, too, has been developed [2,3].
The idea went as follows.

By combining Heisenberg's uncertainty principle with
gravitation, the following relation has been obtained for
the minimum uncertainty ${\Delta}s$ of a single
(timelike) geodesic:
$${\Delta}s^{2}={\alpha}^{4/3}s^{2/3},\eqno(1)$$
where $s$ is the length of the geodesic and $\alpha$ is
the Planck length [c.f. Eq.(3.1) of Ref.1].
Then this uncertainty is believed to be a universal
lower bound, and so
must appear in the space time in an objective way.
This was done via random "gravitational waves".

The present authors [4] reanalysed the concept leading
to Eq.(1). A result is that in Refs.1 and 2 the value
$M$ of mass realizing the least uncertainty along the given
geodesic takes irrealistically high values
$\sim{\hbar\over c}\alpha^{-4/3}s^{1/3}$. For example,
a geodesic of length $s=1$ lightsecond would require a mass
$M\sim10^{10}g$ to be "realized". By other words, the optimum mass
of a clock to measure a period of $1s$ would weigh ten thousand
metric tons.

This result does not directly invalidate the concept
of Refs.1 and 2. Namely, the argumentation needs only
the {\it existence} of a certain lower absolute bound for the
uncertainty; it does not involve {\it real} clocks directly.
However, the high mass problem is intimately connected
with another problem as will immediately be seen.

The original paper [1] as well as the further ones
[2,3] propose that the space-time uncertainties be represented
by {\it random} gravitational waves. These gravitational waves
$\gamma$ satisfy the linearized vacuum Einstein equations:
$$\bar\sqcup\gamma=0\eqno(2)$$
see Eq.(3.2) of Ref.1. Adopting all the time the
conservative notations
of Ref.1, the gravitational wave $\gamma(x,y,z,t)$ is
expanded as a superposition of plane waves:
$$\gamma =\sum_k
\ck cos(k_xx)\cos(k_yy)\cos(k_zz)\cos(kct)+\cdots\eqno(3)$$
The random coefficients $\ck$ are uncorrelated. Their average
is zero while the spreads are given by
$$L^3\bar {\ck^2}=\alpha^{4/3}k^{-5/3}\eqno(4)$$
where $L$ is the normalization volume [c.f. Eqs.(3.4) and
(3.5) of Ref.1]. The above equation is the only one which is
conform to the uncertainty relation (1).

According to the intentions implicit in Refs.1 and 2,
the space-time geometries defined by Eqs.(2) and (3) must
be approximate solutions of the Einstein equations. However,
it turns out that they will not. Though they satisfy,
by construction, the {\it linearized} vacuum Einstein
equations (2), the conditions for the linear approximation
will seriously fail. We are going to test two rather trivial
conditions. The first will hold but the second will not.

Let us calculate the mean squared deviation of the
metric tensor from its Minkowski value. Squaring both
sides of the Eq.(3) and taking stochastic averages of the
coefficients $\ck$, one obtains:
$$\bar{\gamma^2} \sim \sum_k \bar{\ck^2} \sim
 \alpha^{4/3}L^{-3}\sum_kk^{-5/3} \sim (\alpha k_{max})^{4/3}.
\eqno(5)$$
One needs a finite cutoff on $k$ otherwise the amplitude of
the random waves would diverge. K\'arolyh\'azy suggests
$k_{max}=10^{13}cm^{-1}$ and this assures that $\gamma $
is much smaller than the unity. This was the first condition
for applying the linear form (2) of the Einstein equations.

As for the second condition, let us first invoke the
expansion of the scalar curvature $R$ up to the second order
in $\gamma $ [c.f.Ref.5]:
$$R={1\over 2}\bar\sqcup\gamma_{ii}-
    {1\over 2}\gamma_{ij}\bar\sqcup\gamma_{ij}+
    {1\over 4}\gamma_{ij,k}\gamma_{ij,k}+
    {1\over 4}(\gamma_{ij,k}-\gamma_{ik,j})
            (\gamma_{ij,k}-\gamma_{ik,j})+
                                     \cdots.\eqno(6)$$
Now, by substituting the waves (3) into this equation, the
first order term indeed vanishes. The magnitude of
the average of the remaining
terms can be estimated by invoking Eq.(4); one obtains:
$$\bar R \sim \alpha^{4/3}k_{max}^{10/3}.\eqno(7)$$
This curvature is extremely high. Using the previous cutoff
we are led to $\bar R \sim 1cm^{-2}$. So the corresponding
fluctuating
metric is not at all the "extremely small smearing" [1] of
the flat space-time, thought before.

According to the exact Einstein equation
$R={8\pi\alpha^2\over\hbar c}T$.
Hence the curvature (7) would assume
an average energy-density in the order of
$$\bar T \sim \hbar c\alpha^{-2/3}k_{max}^{10/3}.\eqno(8)$$
Observe the dramatic change: in the energy-density the Planck
length appears with
{\it inverse} (two-thirds) power. Therefore the interplay of two
small length scales may result in anything.
The original cutoff $k_{max}=10^{13}cm^{-1}$ would yield
$${\bar T\over c^2}\sim 10^{26}g/cm^3,\eqno(9)$$
i.e. 11 orders of magnitude {\it above neutron star density}.

In Ref.1 the details of the cutoff were thought of no importance.
We have, however, pointed out that the original cutoff would imply
absurd results for cosmological mass density. Since the cutoff
$k_{max}$ is the only free parameter in the model one may hope
to save the theory by choosing a lower value for it. Unfortunately,
the choice $k_{max}=10^5cm^{-1}$, familiar from e.g. the model
of Ghirardi et al. [6], yields still water density. Further decrease
of $k_{max}$ is needed. Then, however, there would be only
macroscopic wavelengths $1/k$ and the gravitational fluctuations (3)
would not play a r\^ole in the quantum-classical transition anymore.
The trace (9) in itself could be removed by means of an incredibly
high cosmologic constant $\Lambda$, but in the Robertson-Walker
Universe geometries two nontrivial components of the Einstein
equations survive, and one cannot remove the problem from both.

Obviously, the K\'arolyh\'azy model [1] has shown to overestimate
something in
the assumed quantum smearing of the space time. The spectrum (4) of
gravitational fluctuations is certainly wrong whatever cutoff is
chosen. The proposals outlined in Refs.[2,3]
derive extremely fine effects to observe experimentally.
In the light of the
cosmological absurdity of the model we wonder if such tiny effects
would have to be taken serious.
\vskip .5cm
The necessity and timeliness to perform the present research
were recognized in a discussion with Prof. P. Gn\"adig of the
E\"otv\"os University. This work was supported by the
Hungarian Scientific Research Fund under Grant No 1822/1991.
\vskip .5cm
\line{\bf References}
\parindent=0cm

[1]\ F.K\'arolyh\'azy, NuovoCim. {\bf XLIIA}, 1506 (1966)

[2]\ F.K\'arolyh\'azy,A.Frenkel and B.Luk\'acs, in:Physics as
natural philosophy, eds. A.Shi\-mony and H.Feschbach (MIT Press,
Cambridge,MA,1982)

[3]\ F.K\'arolyh\'azy,A.Frenkel and B.Luk\'acs, in:Quantum
concepts in space and time, eds. R.Penrose and C.J.Isham
(Clarendon,Oxford,1986)

[4]\ L.Di\'osi and B.Luk\'acs, Phys.Lett. {\bf 142A},331(1989)

[5]\ C.W.Misner,K.S.Thorne,J.A.Wheeler: Gravitation (Freeman,
San Fran\-cisco, 1973)

[6]\ G.C.Ghirardi,A.Rimini and T.Weber, Phys.Rev.{\bf D34},
470 (1986)
\end